\documentstyle[aps,multicol,twocolumn]{revtex}
\begin{document} 
\wideabs{
\title{Why is the black hole entropy (almost) linear in the
    horizon area?}  \author{Gilad Gour\thanks{Electronic mail:
      gour@cc.huji.ac.il} and Avraham E. Mayo\thanks{Electronic mail:
      Mayo@cc.huji.ac.il} } \address{\it The Racah Institute of
    Physics, Hebrew University of Jerusalem,\\ Givat Ram, Jerusalem
    91904, Israel}
 
  \date{\today} \maketitle
 
\begin{abstract} 
  We show that essentially pure classical thermodynamics is sufficient
  to determine Bekenstein's formula for the black hole's entropy,
  $S=\eta A$. We base our reasoning on the minimal assumption that
  since black body radiation is describable by classical
  thermodynamics, so is the complete black hole-Hawking radiation
  system. Furthermore, we argue that any non-linear correction to the
  black hole entropy must be quantum mechanical in nature. The
  proportionality coefficient, $\eta=1/4\ell_P^2$, must be calculated
  within a semi-classical or full-fledged quantum mechanical
  framework.
\end{abstract} 

\bigskip

PACS number(s): 04.70.Dy, 04.70.-s 04.70.Bw

}

\section{Introduction}\label{int}
 
It has been a quarter of a century since Bekenstein \cite{Bek1973} and
Hawking \cite{Haw1975} suggested that the entropy of a black hole is
one fourth of its surface area:
\begin{equation} 
S_{BH}={1\over 4} {A_{\cal H}\over \ell_P^2}. 
\end{equation} 
 
Despite considerable effort \cite{Bek1994} on the quantum
\cite{quantum}, dynamic \cite{dynamic}, and statistical
\cite{statistical} origins of black hole thermodynamics, the exact
source and mechanism of the Bekenstein-Hawking black hole entropy
remain unclear \cite{clear}.  By using the brick wall model, 't Hooft
\cite{breakwall} identified the black hole entropy with the entropy of
a thermal gas of quantum field excitations outside the event horizon,
whereas Frolov and Novikov \cite{quantum} argued that the black hole
entropy can be obtained by identifying the dynamical degrees of
freedom of a black hole with the states of all fields which are
located inside the black hole. A black hole acts as classical
thermodynamic object, but its true microscopic structure remains
unclear \cite{Maggiore1994}.
 
One of classical thermodynamics' aspects is that the entropy and
temperature of many systems can be derived within its framework
without any need for referring to an underlying quantum theory. An
outstanding example is the Stephen-Boltzmann formula for the energy
and entropy flux densities of the radiation emitted by a hot black
body \cite{proof}
\begin{equation} 
u=\sigma T^4, \quad  s={4\over3}\sigma T^3. \label{StBol} 
\end{equation} 
Historically this formula was derived before the discovery of quantum
mechanics.  The coefficient $\sigma=\pi^2/15\hbar^3$ was calculated
only after a quantum mechanical grasp of the phenomenon was achieved.
 
Reapplying this train of thoughts, we wish to establish here that the
proportionality between black hole entropy and horizon area is a
classical result. Any correction to the black hole entropy formula,
$S=\eta A$, must be quantum mechanical in nature {\it i.e.} not
derivable from classical thermodynamics. And just as it happened with
$\sigma$, so here the coefficient $\eta=1/ 4\ell_P^2$, must be
calculated within a semiclassical or a quantum statistical framework
\cite{review}.
 
Such an attempt has been made a few years ago by Gould \cite{Gould},
who claimed to establish the proportionality between black-hole
entropy and horizon area based on the assumption that black holes are
classical thermodynamics objects. After giving evidence for the
relation $S=F(A)$ (to be rediscussed in Sec.~\ref{evidence} below),
Gould uses the principle of equivalence to argue that the {\it local
temperature} of radiation in the exterior of the black hole may depend
only on the {\it local gravitational ``pull''} felt by the local
stationary observer. Accordingly, he concludes that the
proportionality factor must be a universal constant of nature. A
straightforward calculation shows that this proportionality factor is
equal to $8\pi F'(A)$, and therefore that $F(A)\propto A$. However, it
seems that Gould's argument is in not really classical. After all, the
general connection between temperature and acceleration as put in
evidence by Unruh\cite{Unruh}, depends on quantum field theory.

In this work, we do recognize the quantum mechanical foundation from
which Hawking radiation was derived, but also assume that inasmuch as
black body radiation may be investigated using the tools of 19th
century classical thermodynamics, so may Hawking radiation.
 
The structure of the paper is as follows. We begin in
Sec.~\ref{evidence} by reviewing the evidence for the relation
$S=F(A)$, where $F$ is a positive definite and non-decreasing
function. Bekenstein's historical argument for the form of the black
hole entropy, $F(A)\propto A$ has evidently some logical gaps in it;
it very much guesses the form of $F$. In Sec.~\ref{investigation} we
try to bridge this gap. It is shown that classical thermodynamics
decrees the relation $S\sim A^\gamma$, with $\gamma$ a constant. In
Sec.~\ref{constrain}, various physical arguments are used in order to
explore this relation and constrain $\gamma$. The arguments are based
upon various ideas at the base of classical thermodynamics, such as
positive temperature for systems in equilibrium, negative specific
heat for self-gravitating systems, and the generalized second law
(GSL). We find that $\gamma>1/2$. Then, assuming that Hawking
radiation can be described by geometrical optics (short wave
approximation) we prove that $5/6<\gamma\leq 1$. In
Sec.~\ref{conclusion} we present further arguments for why the
relation $S\propto A$ ($\gamma=1$) should be singled out. Finally, the
values of the proportionality coefficient and additive constant are
discussed.
 
Throughout we assume that the system under investigation is embedded
in a $3+1$ spacetime. The question concerning the possible nonlinear
relation between black hole entropy and horizon area in higher
dimensions, was addressed recently by Castro, Graniky and Naschie
\cite{Castro}. Furthermore, we deal only with bare black holes,
excluding the so called "dirty" black holes, which are known to have a
modified entropy formula due to the presence of classical fields on
the horizon \cite{Visser}.
 
We use units with $\hbar=G=c=k_B=1$, unless otherwise stated.
 
\section{Evidence for $S=F(A)$}\label{evidence}
 
Our framework is classical thermodynamics. We shall not use quantum
mechanics nor statistical mechanics. The black hole and surroundings
are treated as a thermodynamical system in equilibrium with definite
temperature $T$ and entropy $S$, which are quantities assumed to be
definite functions of the black hole macroscopic parameters. In the
spirit of the ``no hair" theorem, these are the mass $M$, charge $Q$
and angular momentum $J$ of the black hole.
 
The {\it first law of thermodynamics} applied to the system black
hole--surrounding may be written as
\begin{equation} 
dM=TdS-d\mathcal{W}, 
 \label{flaw} 
\end{equation} 
where $\mathcal{W}$ is the work done on (or extracted from) the black
hole. Specifically, let $\Phi$ and $\Omega$ be the electric potential
and the angular velocity on the event horizon respectively. Then the
work done (energy extracted) due to changes in angular momentum and
electric charge is
\begin{equation} 
{d\mathcal{W}}=\Phi dQ+\Omega dJ. 
\end{equation} 
Substituting this in Eq.~(\ref{flaw}) gives
\begin{equation} 
TdS=dM-\Phi dQ-\Omega dJ. \label{ent} 
\end{equation} 
By exploring the dynamics of test objects in the black hole exterior,
$\Phi$ and $\Omega$ are found to be \cite{MTW}
\begin{equation} 
\Phi={4\pi r_+Q\over A},\quad \Omega={4\pi a\over A} \label{phio} 
\end{equation} 
where $r_+=M+\sqrt{M^2-Q^2-a^2}$ and $A=4\pi(r_+^2+a^2)$ are the
radius and area of the event horizon respectively, and $a\equiv J/M$
is the specific angular momentum.
 
We proceed by differentiating the horizon area $A$ with respect to
$M$, $Q$ and $J$. The result is nothing but the {\it first law of
  black holes mechanics}:
\begin{equation} 
{1\over 8\pi}\kappa dA=dM-\Phi dQ-\Omega dJ \label{area}, 
\end{equation} 
where $\kappa$ is the surface gravity given by
\begin{equation} 
\kappa=8\pi\left({\partial M\over\partial A}\right)_{J,Q} 
={4\pi\sqrt{M^{2}-Q^{2}-a^{2}}\over A}. \label{kappa} 
\end{equation} 
 
Combining now Eqs.~(\ref{ent}) and (\ref{area}) gives:
\begin{equation} 
  {1\over 8\pi}\kappa dA=TdS, 
\label{dads}
\end{equation} 
which impels us to infer that {\it the black hole entropy must be a
  definite function of its horizon area}:
\begin{equation} 
S=F(A). 
\end{equation} 
Gould \cite{Gould} obtained the same result based on the argument that
two black holes with the same horizon area must have the same entropy,
since otherwise one would be able to violate the GSL by Penrose
processes \cite{Penrose}.

An interesting question is whether the result that $S=F(A)$ is
generic, namely, does it apply to others types of black holes besides
the Kerr-Newmann black hole? The answer to this query seems to be on
the affirmative. For example, a $(2+1)$-dimensional (BTZ) black hole
obeys the standard {\it first law of thermodynamics} (\ref{ent})
augmented by an additional work term, $-{\cal P} d(2\pi R)$, where ${\cal
P}$ is the surface pressure at the boundary of a cavity of radius
$R$. Accordingly, the same arguments which served to prove that the
relation $S=F(A)$ for the Kerr-Newman black hole, may be used now to
prove that the entropy of the BTZ black hole must be some function of
its horizon area, and indeed, semi-classical calculations yields a
linear entropy to horizon area relation \cite{Carlip}. Another example
is the dilatonic black hole for which the entropy is proportional to
the horizon area \cite{Mann}. Keeping these other interesting cases in
mind, we limit our analysis to the case of the Kerr-Newman black hole,
since it captures the essentials of the ideas discussed in this work.
 
The result $S=F(A)$ clashes with a recent result by Vaz and
Witten\cite{Vaz}, who showed that in the framework of canonical
quantum gravity the entropy of a charged black hole turns out to be
the difference between the outer and inner horizon areas. Vaz and
Witten explain this disagreement with the semi-classical result, as
the product of neglecting the effect of back reaction to any radiation
emitted exterior to the black hole. Another explanation involves a
similar effect induced by emission of radiation in the interior of the
black hole, radiation which is undetectable by an exterior observer.
However, as Vaz and Witten admit themselves, their result may be due
to the too restrictive boundary conditions imposed on the wave
functional in the interior of the black hole (the wave functional is
made to vanish beyond the inner horizon). However, they hedge this
claim by pointing out that other boundary conditions are in effect
unknown.
 
Keeping this in mind we proceed by noting that the black hole
temperature is given by
\begin{equation} 
T=\left({\partial M\over \partial S}\right)={\kappa\over 8\pi 
F'(A)}={\sqrt{M^{2}-Q^{2}-a^{2}}\over 2AF'(A)}. \label{temp} 
\end{equation} 
 
Interestingly, for the black hole to have a non negative temperature,
$F'(A)>0$ {\it i.e.} $F(A)$ must be a monotonic non-decreasing
function. In view of Hawking's increasing area theorem, this result
had been expected, and indeed was assumed by Bekenstein\cite
{Bek1973}. In the rest of the paper we resort to classical
thermodynamics to specify $F(A)$.
 
\section{Investigating $F(A)$}\label{investigation}
 
Leaning on the parallelism between the {\it zeroth law of black hole
  mechanics} ($\kappa$ is constant over the whole of the event horizon
surface of a stationary black hole), and the {\it zeroth law of
  thermodynamics} ($T$ is uniform over a system in equilibrium), one
might conclude that the thermodynamical temperature must be a definite
function of the surface gravity, and then invoke Eq.~(\ref{temp}) to
conclude that $F'(A)=const$, $F$ is linear in $A$, and the proof is
complete.
 
However, there is a loophole: $T$ may also depend on $\Phi$ and
$\Omega$, which are themselves constants on the horizon. Hence, it is
less clear why $F$ should be linear in $A$. In order to go forward, we
first prove the following statement:
 
\smallskip
 
{\it The black hole entropy must have the form:
\begin{equation} 
S=\eta A^{\gamma}+S_0, \label{lem1} 
\end{equation} 
where $\eta$ and $S_0$ are constants of integration and $\gamma$ is a
dimensionless parameter.}
 
\smallskip
 
To prove this, we consider the Reissner-Nordstrom black hole (the
derivation based on the Kerr solution is similar). Taking the
logarithm in both sides of Eq.~(\ref{temp}) and differentiating gives
\begin{equation} 
{dT\over T}+\left(A{F''(A)\over F'(A)}+1\right){dA\over A} ={MdM-QdQ\over 
M^2-Q^2}. 
\end{equation} 
Consider now an isothermal process ($dT=0$). Then
\begin{equation} 
1+A{F''(A)\over F'(A)}={A\over M^2-Q^2}{MdM-QdQ\over dA}\mid _{T={\rm 
const.}} \label{expf} 
\end{equation} 
Using now Eq.~(\ref{area}) with $\Omega=0$, we find that
\begin{equation} 
{MdM-QdQ \over dA}\mid _{T={\rm const.}}={\kappa \over 
8\pi}{M\left({\partial M \over \partial Q}\right)_T-Q \over 
\left({\partial M \over \partial Q}\right)_T-\Phi}. 
\end{equation} 
Substituting this and the expressions for $\Phi$ and $\kappa$
(Eqs.~(\ref{phio}) and (\ref{kappa}), respectively) in
Eq.~(\ref{expf}), and using the dimensionless parameter $y\equiv Q/M$,
we find that
\begin{eqnarray} 
1+A{F''(A)\over F'(A)} &=&{1\over 2\sqrt{1-y^2}}{\Phi_T-y\over 
\Phi_T-{y\over 1+\sqrt{1-y^2}}}, \nonumber \\ 
\Phi_T&\equiv&\left({\partial M\over \partial 
Q}\right)_T. \label{final} 
\end{eqnarray} 
 
Keep in mind that $|y|\leq 1$, where equality is achieved for extremal
black holes. $\Phi_T$ can be interpreted as the electric potential on
the horizon when the black hole is in equilibrium with a surrounding
heat bath. Since we are interested in the classical regime, we expand
$\Phi_T$ in powers of $\hbar$ with the leading term, $O(\hbar^0)$ is
taken to be classical. Higher order terms in $\hbar$ are considered to
be quantum corrections. The classical leading term in $\Phi_T$ cannot
vanish because in the limit $y\rightarrow 1$ the right hand side of
Eq.~(\ref{final}) diverges unless $\Phi_T\rightarrow 1$. This may be
anticipated, because in the limit of zero temperature (that is $y=1$)
$\Phi_T$ should be equal to $\Phi$ because the black hole can be
considered to be isolated from the thermal bath ({\it i.e.} $T=0$).
Hence, $\Phi_T(y=1)=\Phi(y=1)=1$.  Accordingly, hereafter we assume
that $\Phi_T$ has a non vanishing classical part.
 
Since $\Phi_T$ is a dimensionless quantity it must be a function of
$y$ only.  Consequently we have achieved separation of the variables
in the problem, $y$ and $A$: the right hand side of Eq.~(\ref{final})
is a function of $y$ only whereas the left hand side is a function of
$A$ only. Since, $A$ and $y$ are two independent parameters this
implies that
\begin{equation} 
1+A{F''(A) \over F'(A)}={\rm constant}\equiv\gamma.\label{fa} 
\end{equation} 
Solving this for $F(A)$ with $\gamma\neq 0$ reproduce
Eq.~(\ref{lem1}).  One must add that for $\gamma=0$, the solution is
logarithmic in $A$. However in the next section we explain why this
possibility must be excluded.
 
\section{Constraining $\gamma$.} \label{constrain}
 
Here we use various arguments to constrain $\gamma$. In subsection
\ref{gamma>1/2} we use some basic principles of classical
thermodynamics to set the constrain $\gamma>1/2$. This result is of
great importance since it emphasizes the fact that black hole
thermodynamics is not like ordinary thermodynamics. Surely the choice
$\gamma =1/2$, {\it e.g.} $S\sim \sqrt{A}$, seems to stands out
because it implies that for $Q=J=0$, $S\propto M$, in harmony with
the extensive character of entropy in ordinary thermodynamics. But as
proven below, this later choice clashes with basic principles of black
holes and traditional physics, and thus should be rejected.
 
Consider now, Hawking radiation. Obviously, Hawking's celebrated
derivation of the black hole radiation is quantum-mechanical.
Consequently it could be argued that by considering this phenomenon we
are actually invalidating our claim that the derivation of black hole
entropy may be based on purely classical arguments. But in a sense
thermodynamics is sensitive to effects at the quantum level; while
classical mechanics is consistent with the limit $\hbar\rightarrow 0$,
classical thermodynamics is not. Accordingly we make the minimal
assumption that the laws of thermodynamics apply to the black hole and
radiation as a complete system (after all, if the black hole is to be
considered a thermodynamical object with entropy and temperature, it
must be able to radiate as well as absorb). This implies that the
radiation may be described by the Stephan-Boltzmann law. As explained
in subsection \ref{rest} this leads us to conclude that
$5/6<\gamma\leq 1$.
 
It should be pointed out that in his article, Gould does not address
the issue of Hawking radiation and the impact it may have on the
derivation of the black hole entropy formula. Hawking himself
discusses it in \cite{Haw1975}.
 
\subsection{Positive temperature, Negative Specific Heat and 
the GSL: $\gamma>1/2$.}\label{gamma>1/2}
 
Consider first the temperature of the Schwarzschild black hole
Eq.~(\ref{temp}),
\begin{equation} 
T={M \over 2AF'(A)}={1 \over 2\eta\gamma (16\pi)^{\gamma}M^{2\gamma-1}}. 
\end{equation} 
The first obvious fact is that $\gamma$ must be positive definite for
the system to have positive definite temperature. Examining next the
specific heat $(\partial T/\partial M)^{-1}\propto 1-2\gamma$; we find
that for the system to have a negative specific heat, as befits a
self-gravitating system, $\gamma$ must be greater then $1/2$. Note
that if the logarithmic solution of Eq.~(\ref{fa}) were to be taken
seriously, it would imply that the black hole temperature is linear in
the mass, and hence the specific heat would be constant independent of
the black hole mass.
 
A somewhat related argument supporting the constraint $\gamma>1/2$,
involves the interplay between the gravitational and the electric
forces in the physics of a Reissner-Nordstrom black hole. It is
reasonable to assume that when the black hole charge is small, the
thermodynamics of the system is governed primarily by gravitation,
which decrees a {\it negative} specific heat. However, as the
magnitude of the black hole charge is increased, electrodynamics
becomes important. At some point the system should begin to show
typical qualities of a traditional thermodynamical system, such as a
{\it positive} specific heat.  Consequently, at some critical values
of $Q$ and $M$, the specific heat should switch sign from negative to
positive through a second order phase transition. For example, for
$\gamma=1$ this transition takes place when $Q=\sqrt{3/4}M$. For a
general $\gamma$ this condition translates to
$Q=\sqrt{1-1/(4\gamma^2)}M$. If this phenomenon is to be reproduced,
$\gamma$ must again be greater than $1/2$.
 
A more rigorous proof for $\gamma>1/2$ can be drawn from the following
{\it gedanken} experiment. Consider two identical Schwarzschild black
holes of mass $M$, which collide and merge to form a third black hole
of mass ${\cal M}$. The initial entropy of the system is
$S_i=2\eta(16\pi M^2)^{\gamma}+2S_0$. During the process of
in-spiraling and finally coalescence, the system lose energy primarily
by emission of gravitational waves and negligibly by thermal
radiation.  Accordingly the mass of the black hole at the end state of
the system must be smaller then the initial energy contained in the
two well separated black holes: ${\cal M}<2M$. It follows that the
entropy at the end state of the system, $S_f$ is bounded from above by
$\eta(64\pi M^2)^{\gamma}+S_0$. But by the GSL it must also be greater
then the entropy at the initial state of the system. Combining these
two arguments gives
\begin{equation} 
2\eta(16\pi M^2)^{\gamma}+2S_0<S_f<\eta(64\pi M^2)^{\gamma}+S_0. 
\end{equation} 
The only way in which this can be true for arbitrary $M$ is for
\begin{equation} 
\gamma>1/2. 
\end{equation} 
 
As a last remark we point out that this {\it gedanken} experiment also
provide us with clear cut evidence against $F(A)=\ln {A}$, since it
implies that the GSL would be violated for $A>4\exp{(-S_0/\eta)}$
(note that here $S_0$ and $\eta$ are both dimensionless).
 
\subsection{Hawking radiation: $5/6<\gamma\leq 1$.} \label{rest}
 
Assume now that the radiation emitted by a Schwarzschild black hole
follows the Stephan-Boltzmann law ~(\ref{StBol}). We are clearly
assuming that geometrical optics may be applied here, that is, that
the characteristic wavelength of the radiation $\lambda_{max}$, is
smaller than the characteristic length scale of the emitter, namely
$2M$. In the regime where $\lambda_{max}\gg M$ geometrical optics
ceases to provide a good description for the radiation, and wave
optics must be employed instead. However, since $\gamma$ is a constant
of the theory, any result concerning $\gamma$ which was obtained
within a regime describable by geometrical optics, should be
extendable to regimes where geometrical optics fails.
 
Keeping this point in mind, we recall that by the principle of black
body radiation, the radiated power (which equals minus the rate of
change of the black hole mass) is given by
\begin{equation} 
\dot{M}\propto-T^4 A\propto- M^{2(3-4\gamma)}. \label{roch} 
\end{equation} 
In the limit $M\rightarrow\infty$, the temperature approaches zero and
thus should also the radiated power $\dot{M}$. This indicate that
$\gamma$ must be bigger than $3/4$. Furthermore, the rate of change of
the black hole entropy is
\begin{equation} 
\dot{S}={\dot{M}\over T}\propto- M^{5-6\gamma}. \label{rofe} 
\end{equation} 
In the limit where $T\rightarrow 0$ ($M\rightarrow\infty$),
$\dot{S}\rightarrow 0$, which implies that $\gamma>5/6$. Also,
combining Eqs.~(\ref{rofe}) and (\ref{roch}), one finds that
\begin{equation} 
\dot{S}\propto\dot{M}^{{5-6\gamma\over 2(3-4\gamma)}}. 
\end{equation} 
It is reasonable to assume that $\dot{S}$ and $\dot{M}$ increase
(decrease) simultaneously. Accordingly, we infer that
$(5-6\gamma)/(3-4\gamma)>0$. Since it was already shown that
$\gamma>3/4$, the constraint $\gamma>5/6$ is reproduced.
 
Consider next Wein's law which teaches us that
\begin{equation} 
\lambda_{max} T\sim\hbar, 
\end{equation} 
where $\lambda_{max}$ corresponds to the wavelength at which the
radiation intensity is maximized. This suggests that
$\lambda_{max}\sim 1/T\sim M^{2\gamma-1}$. Recalling now that this
would be true provided that
\begin{equation} 
{\lambda_{max}\over 2M}\sim M^{2(\gamma-1)}\lesssim 1, 
\end{equation} 
we accept that $\gamma\leq 1$. Thus we conclude that
\begin{equation} 
5/6<\gamma\leq 1. 
\end{equation} 
 
\section{Arguments for $\gamma=1$. What can be said about $S_0$ 
  and $\eta$?}\label{conclusion}
 
The first argument is based on the observation that the power by which
the temperature is raised in the Stephan-Boltzman law is determined by
the effective dimension of the phase space by which the system is
described. The rule that seems to emerge is that the power by which
the temperature is raised is equal to ${\cal D}+1$, where ${\cal D}$
is the effective dimension of the system. Keeping this in mind, we
rewrite the expression for the radiated power (\ref{roch}) in term of
the temperature
\begin{equation} 
\dot{M}\propto-T^4 A\propto-T^{4- {2\over 2\gamma-1}}. 
\end{equation} 
This suggests that the black hole is in effect a thermodynamic system
with an effective dimension ${\cal D}$, fixed by the equality
\begin{equation} 
{\cal D}+1=4-{2\over 2\gamma-1}.\label{red1} 
\end{equation} 
The constraint on $\gamma$ may be used now to constrain ${\cal D}$ and
even to tell us something about $\gamma$. Solving Eq.~(\ref{red1}) for
$\gamma$ we find
\begin{equation} 
\gamma={1\over 2}+{1\over 3-{\cal D}}. 
\end{equation} 
Using the result of the previous section, $5/6<\gamma\leq1$, we infer
that $0<{\cal D}\leq 1$. Assuming that ${\cal D}$ is an integer
(glossing over the possibility of fractal dimensions which is usually
an attribute of systems lacking any kind of characteristics scale), we
find that $\gamma$ may have only one possible value,
\begin{equation} 
\gamma=1. 
\end{equation} 
Remarkably, even though the black hole is evidently a $3D$ object
embedded in a $3+1$ dimensional spacetime, it behaves as if it was a
{\it one dimensional} thermodynamical object!
 
Imagine now a number of black holes in vacuum momentarily at rest at
some distance one from another \cite{Bek2000}. The only evident source
of entropy is the black holes horizons. Presuming that the entropies
of independent systems are additive, we may write
\begin{equation} 
S_{total}=\sum_i (\eta A_i^\gamma+S_0). \label{st} 
\end{equation} 
We make the assumption that the entropy contributed by a specific
black hole is not changed by its motion and associated dynamical
changes, induced by its companions, so that the formula for
$S_{total}$ applies also when the black holes falls towards each
other. Focus now on two of these black holes, with areas $A_1$ and
$A_2$, as they fall towards each other and merge into a single one. By
Hawking's result each $A_i$ is bound to increase\cite{Haw71}.
Furthermore, the growth of the {\it event horizon} area is a
continuous process (to be contrasted with the growth of the {\it
  apparent horizon} area which may be discontinues).  Therefore, at
the moment of coalescence the horizon area of the new black hole is
the sum of the areas of the merging constitutes, $A_{new}=A_1+A_2$. By
the assumption that the relation between entropy and area is still
valid in the case of distorted black holes, we may use Eq.~(\ref{st})
for the new black hole.  During the infall and the merger processes,
semi-coherent gravitational radiation is emitted, which should not
have much effect on the entropy balance. Hence at the moment of
merger,
\begin{equation} 
(A_1+A_2)^\gamma= A_1^\gamma+A_2^\gamma+S_0/\eta. 
\end{equation} 
Obviously one can arrange for the process to occur with $A_1=A_2=A_0$,
for example by slowly lowering two black holes of equal masses towards
each other, so that the area increase and distortion are identical in
both black holes. This must be true for any arbitrary $A_0$. So we are
led to the conclusion that $\gamma=1$ and $S_0=0$. This agree with the
reasonable assumption that the entropy tends to zero as the mass tends
to zero. Moreover, the vanishing of the additive coefficient $S_0$
indicates that the zero of the entropy is set, without the liberty of
adding a constant, as one can do in classical physics.
 
What can be said about the proportionality coefficient $\eta$?
Obviously, $\eta$ should have dimensions of inverse area. Further
insight seems to require the use of quantum physics. However,
Wheeler's heuristic suggestion that the right order of magnitude of
$S$ should be gotten by just dividing $A$ by the Planck length squared
$\ell_P^2$, gained support from the observation by Bekenstein
\cite{Bek1973} that when an elementary particle is very softly
deposited at the horizon of {\it any} Kerr-Newmann black hole, the
minimal increase of $S$ so calibrated is of order unity. Since an
elementary particle should carry no more than a unit of entropy, the
GSL would fail if we took $\eta$ as an inverse length square with a
length large on the Planck scale. Also it makes no sense to take this
length scale to be smaller than Planck's length, which is regarded as
the smallest scale on which smooth spacetime is a reasonable paradigm.
Nothing could be said about the numerical magnitude of
$\eta_0\equiv\eta \,\ell_P^2$ without over-stepping the realm of
classical thermodynamics.
 
Thus we close by writing
\begin{equation} 
S_{BH}=\eta_0 {A_{\cal H}\over \ell_P^2}. 
\end{equation} 
 
\medskip
 
{\bf ACKNOWLEDGMENTS} It is a pleasure to thank Prof. J. D. Bekenstein
for suggesting this problem and for his guidance during this work.
This research is supported by a grant from the Israel Science
Foundation, established by the Israel Academy of Sciences and
Humanities.


\begin{references}
\bibitem{proof} The textbook proof goes as follows. Let us make a
  blackbody enclosure with a piston, so that work may be done on or
  extracted from the radiation. By the first law of thermodynamics, we
  have $dQ=dU+pdV$, where $Q$ is heat and $U$ is total energy. By the
  second law of thermodynamics $dS=dQ/T$, where $S$ is the entropy.
  Since $U=u(T)V$, and $p=u(T)/3$, we have $dS=(V/T)u'(T)
  dT+(4u(T)/3T)dV$. Further, $dS$ is a perfect differential, thus we
  obtain {\small $${1\over T}u'(T)={\partial^2 S\over \partial
      T\partial V}=-{4u(t)\over 3T^2}+{4\over 3T} u'(T),$$} so that
  $u'(t)=4u(T)/T$ and thus $u(T)=aT^4$, where $a$ is a constant of
  integration. Thus we obtain the Stephan-Boltzmann law.
\end{references}
\end{document}